\documentclass[letterpaper, preprint, paper,11pt]{AAS}	

\usepackage{bm}
\usepackage{amsmath}
\usepackage{mathabx}
\usepackage{subfigure}
\usepackage{tabularx}
\usepackage{booktabs}
\usepackage[colorlinks=true, pdfstartview=FitV, linkcolor=black, citecolor= black, urlcolor= black]{hyperref}
\usepackage{overcite}
\usepackage{footnpag}			      	
\usepackage{caption}
\usepackage{comment}

\PaperNumber{21-785}

\begin{document}

\title{Gravity Estimation at Small Bodies via Optical Tracking of Hopping Artificial Probes}

\author{Jacopo Villa\thanks{Graduate Student, Aerospace Engineering Sciences, University of Colorado Boulder, 3775 Discovery Drive, Boulder, CO 80303}, Andrew French\thanks{Navigation Engineer, Jet Propulsion Laboratory, California Institute of Technology, 4800 Oak Grove Drive, Pasadena, CA 91109, USA}, Jay McMahon\thanks{Assistant Professor, Aerospace Engineering Sciences, University of Colorado Boulder, 3775 Discovery Drive, Boulder, CO 80303}, Daniel Scheeres\thanks{Distinguished Professor, Aerospace Engineering Sciences, University of Colorado Boulder, 3775 Discovery Drive, Boulder, CO 80303}, \ and Benjamin Hockman\thanks{Robotics Technologist, Jet Propulsion Laboratory, California Institute of Technology, 4800 Oak Grove Drive, Pasadena, CA 91109, USA}
}

\maketitle{}

\begin{abstract}
Despite numerous successful missions to small celestial bodies, the gravity field of such targets has been poorly characterized so far. Gravity estimates can be used to infer the internal structure and composition of small bodies and, as such, have strong implications in the fields of planetary science, planetary defense, and in-situ resource utilization. Current gravimetry techniques at small bodies mostly rely on tracking the spacecraft orbital motion, where the gravity observability is low. To date, only lower-degree and order spherical harmonics of small-body gravity fields could be resolved. In this paper, we evaluate gravimetry performance for a novel mission architecture where artificial probes repeatedly hop across the surface of the small body and perform low-altitude, suborbital arcs. Such probes are tracked using optical measurements from the mothership's onboard camera and orbit determination is performed to estimate the probe trajectories, the small body's rotational kinematics, and the gravity field. The suborbital motion of the probes provides dense observations at low altitude, where the gravity signal is stronger. We assess the impact of observation parameters and mission duration on gravity observability. Results suggest that the gravitational spherical harmonics of a small body with the same mass as the asteroid Bennu, can be observed at least up to degree 40 within months of observations. Measurement precision and frequency are key to achieve high-performance gravimetry.
\end{abstract}

\section{Introduction}

Our understanding of primitive bodies in the Solar System has grown significantly over the last two decades, especially thanks to an ever increasing number of exploration missions. Just recently, NASA's OSIRIS-REx and JAXA's Hayabusa2 collected a wealth of scientific data around the Near-Earth Asteroids Bennu and Ryugu, respectively, and successfully sampled surface material to be studied on Earth. Several missions toward small bodies are planned for the near future: NASA’s Janus, Psyche, and Lucy will explore the exotic surroundings of binary asteroids, a nickel-iron small body, and Main Belt’s as well as Trojan asteroids, respectively; AIDA, a NASA/ESA joint mission, will test and assess asteroid deflection strategies in orbit for the first time. Whilst past missions provided invaluable insights on the geophysics of these worlds, many questions about small bodies remain unanswered. One key feature requiring further investigation is the interior structure and composition. Evidence shows that small body interiors can vary dramatically, from mostly homogeneous to largely heterogeneous composition, with under- or over-dense regions, including voids typical of rubble piles. However, there is a lack of direct observations to confirm current geophysical models\cite{scheeres2015asteroid}. Enhancing the knowledge of small body interiors could help to (1) validate and refine current models of the solar system formation and evolution, (2) assess the effectiveness of asteroid deflection techniques, and (3) augment subsurface situational awareness for asteroid mining and resource exploitation purposes\cite{michel2015asteroids}. 

Three main techniques have been used and proposed to map the interiors of small bodies: radar tomography, seismic imaging, and gravimetry. Radar tomography can characterize the 3D interior structure of a small body by measuring the reflection and transmission of ground-penetrating radio signals. This technique has been successfully employed to measure the dielectric constant of comet 67P during the Rosetta mission and several other mission concepts based on radar tomography have been proposed\cite{ciarletti2015consert}. However, it requires more complex mission architectures and instrumentation, and presents multiple challenges for the global mapping of a small body. First, 3D reconstruction of the deeper areas can be difficult, due to signal attenuation and unfavorable observation geometry. Second, data inversion can be challenging and unreliable, and it is recommended that this technique is informed by other interior-mapping strategies, such as gravimetry. Third, certain materials composing small bodies can be opaque to electromagnetic energy, undermining the effectiveness of this method. Finally, overall performance strongly depends on the target body's dielectric properties as well as the selected mission architecture and instrumentation, requiring mission-specific design and pre-mission calibration\cite{sorsa2019bistatic,haynes2020asteroids}. 

Seismic imaging measures the elastic waves traversing the body to infer its interior properties, and has been widely used to map the internal structure of the Earth. Several mission concepts have been proposed to apply the same principle at small bodies, deploying seismometers and detonators on the surface\cite{scheeres2003asteroid,plescia2017asteroid}. However, a number of practical challenges arise, such as the precise deployment and localization of surface probes and the unknown dissipation properties of the target body. A simpler technique to infer crude estimates of a small body's interior properties consists in artificially inducing seismic shaking with a kinetic impactor. Observing the displacement of surface features caused by the impact event can reveal some of the body's internal properties, such as the bulk density. This technique was recently adopted by the Hayabusa2 spacecraft at the asteroid Ryugu; however, no relevant regolith displacements were detected, possibly due to the porosity and plastic behavior of the asteroid's surface, which dissipated seismic waves caused by the impact event\cite{nishiyama2021simulation}. 

Finally, gravity measurements (or gravimetry) can be used to interpret the interior properties of a small body, in particular, its mass distribution. Gravimetry has a long history of successes in small body missions, and consists in tracking the spacecraft motion (usually with ground-based radiometric measurements and onboard optical measurements) in the vicinity of the small body, co-estimating the spacecraft trajectory and the central body's gravity field. Since this technique is inherently part of the navigation process, it does not necessarily require additional instruments and operations. Ideally, a high-precision estimate of the gravity field would require low-altitude observations (where the gravity signal is stronger), large spatial distribution of measurements around the body and precise measurements. This is especially important for low-mass target bodies, as the gravity observability is lower. However, mission requirements usually constrain spacecraft to fly far from the surface, where orbits are more stable, but the gravity signal is weaker. As such, the low-signal gravity anomalies remain undetected and, in turn, the heterogeneous mass distribution largely undiscovered. 

To date, gravity estimation based on spacecraft tracking only provided estimates for the lower degrees and orders of the gravity field spherical harmonics, which in turn only permits low-fidelity characterization of the interior mass distribution. On the other hand, the OSIRIS-REx mission recently pioneered a novel gravimetry approach: the mission team coincidentally discovered natural ejecta flying in the vicinity of the asteroid Bennu and, by tracking the particles motion, could estimate the asteroid's gravity field to about degree and order 8 \cite{chesley2020trajectory}. For comparison, only the $J_2$ coefficient of Bennu was recovered by tracking the spacecraft, at the end of the Orbital A phase of that same mission\cite{leonard2019osiris}. This unprecedented success proved the potential of measuring low-altitude, spatially-distributed objects to drastically improve gravimetry performance at small bodies. However, tracking natural ejecta comes with limitations: these natural events are not controllable nor predictable, and they may not occur at all when visiting a given small body. Furthermore, the ejecta are fragments of a low-albedo celestial body and their size in the camera plane is sub-pixel. Hence, they are associated with large astrometric uncertainty which, in the OSIRIS-REx case, with a median of 1.05 pixels\cite{chesley2020trajectory}. Additionally, such particles can hardly be detected in front of the small body, since their photometric properties match the one of the body surface. As such, a large fraction of low-altitude measurements (those carrying the stronger gravity signal) cannot be detected and used for gravimetry.

\begin{figure}[]
	\centering\includegraphics[width=0.6\textwidth]{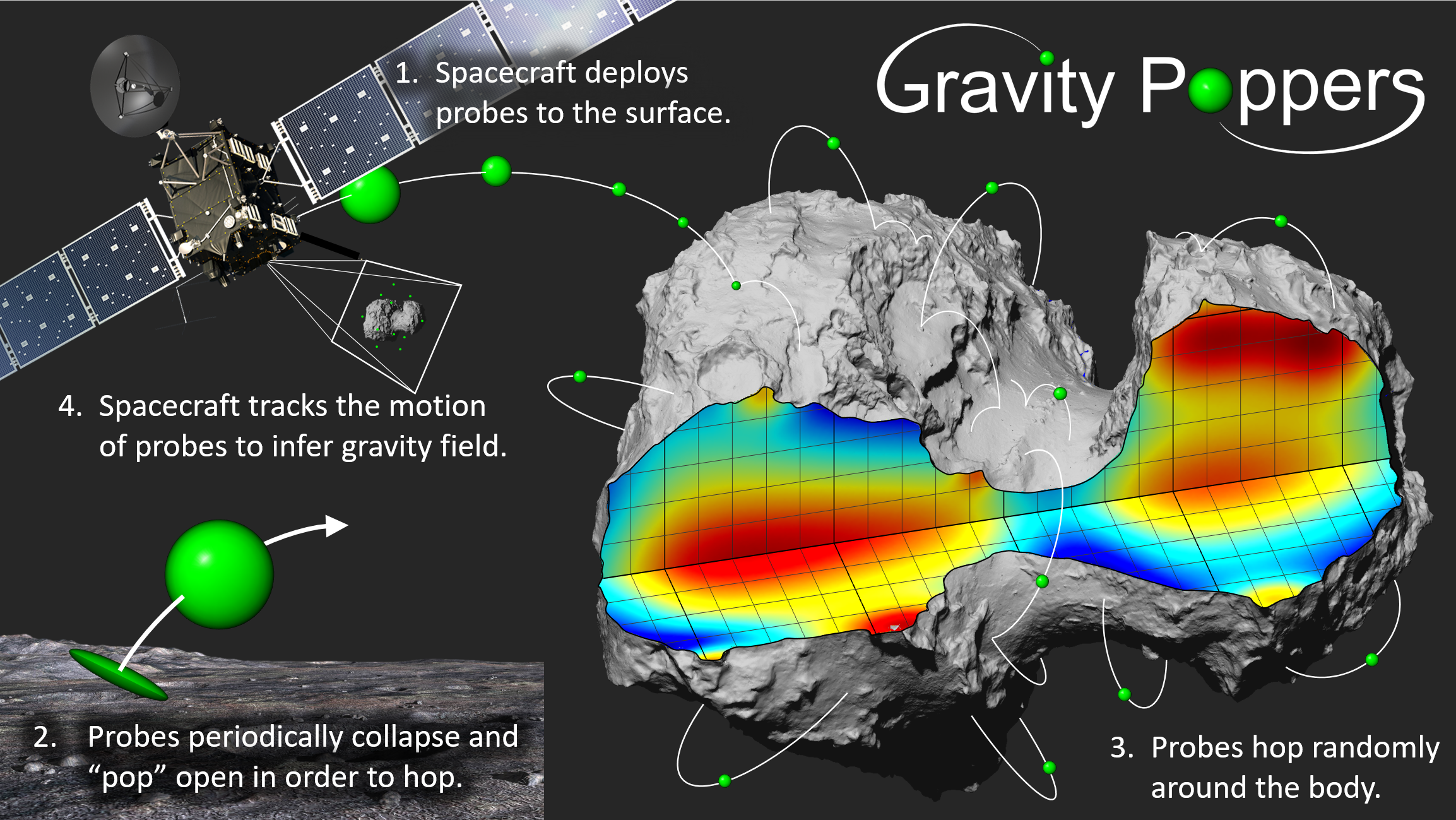}
	\caption{Infographic of the Gravity Poppers mission concept.}
	\label{fig:ConceptFig}
\end{figure}

In this paper, we evaluate gravimetry performance for a novel mission concept that leverages measurements of low-altitude, spatially-distributed objects for gravimetry at small bodies. Instead of relying on serendipitous and unpredictable observations of natural particles, we propose to track a swarm of artificial probes designed to maximize gravimetry performance. We assume that such probes are characterized by two key features: (i) they can repeatedly hop from the surface of the small body, performing suborbital motion, and (ii) they are visible both on the dark background and in front of the small body surface, from the observer's viewpoint. The probes motion is observed by a mother spacecraft hovering in the vicinity of the small body. This mission architecture was firstly presented as a NASA Innovative Advance Concept (NIAC) study, named \textit{Gravity Poppers}, by Hockman et al. and is proposed as a low-budget approach for high-precision gravimetry at small bodies\footnote{\url{https://www.nasa.gov/directorates/spacetech/niac/2020_Phase_I_Phase_II/Gravity_Poppers/}}. An infographic of this concept is shown in Figure \ref{fig:ConceptFig}. In its simplest design, this concept only requires a camera aboard the mothership to observe the probes motion. The probes could be actively emitting light with LEDs or passively reflecting the sunlight. This paper is structured as follows: we firstly define the problem statement and the simulation setup. Second, we present the assumptions on the dynamical environment and measurements. Third, we present orbit determination (OD) results and how they can inform the mission design.

\section{Problem Statement}

The following mission architecture is considered: a set of artificial probes performs suborbital motion around a small body, repeatedly hopping from its surface once they come at rest. The probes motion is observed by a mother spacecraft, hovering in the vicinity of the small body, which detects the probes using the onboard camera. Using optical measurements, the gravity field of the small body is estimated using an OD process. The objective of this work is threefold: (1) quantify gravity estimation performance for a notional and representative mission design, (2) analyze the sensitivity of gravimetry performance to the most relevant design parameters, and (3) present order-of-magnitude requirements for mission duration, measurement performance, and mission architecture.

\section{Simulation Setup}

This study is carried out using numerical simulations. The simulated scenarios are implemented using the Mission Analysis, Operations, and Navigation Toolkit Environment (MONTE) developed by NASA Jet Propulsion Laboratory (JPL)\cite{evans2018monte}. MONTE provides a state-of-the-art toolkit for mission analysis, trajectory design, flight path control, and navigation and is used as the prime orbit determination software for missions flown by JPL. MONTE can be used as an importable Python module and we integrate it in a broader Python code for post-processing and data analysis.

\begin{figure}[]
	\centering\includegraphics[width=0.49\textwidth]{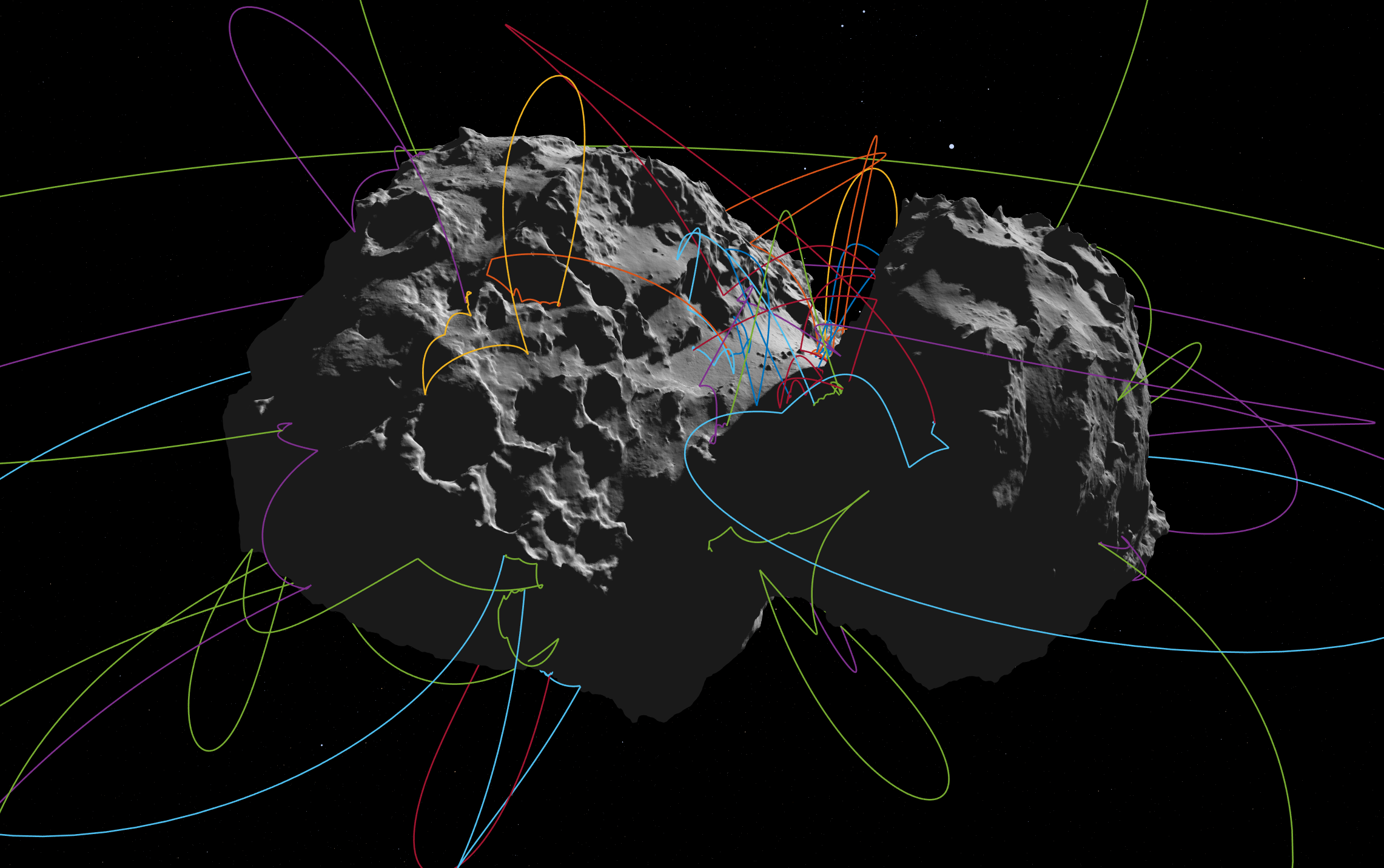}
	\centering\includegraphics[width=0.49\textwidth]{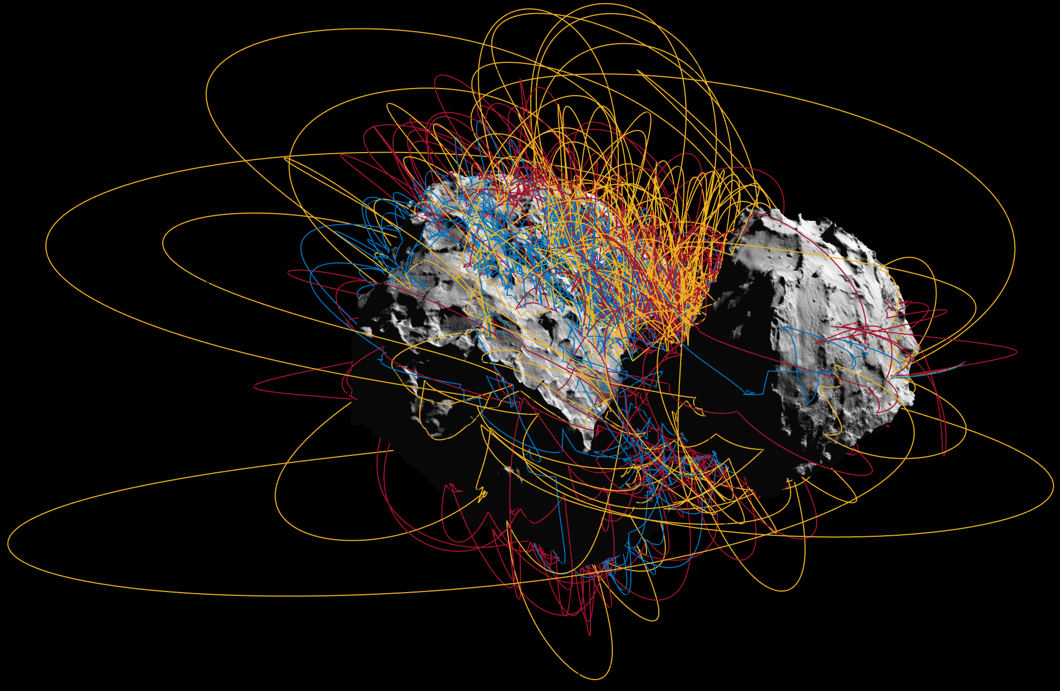}
	\caption{Simulated trajectories of hopping probes around comet 67P Churyumov-Gerasimenko, in the small body-fixed frame. The dynamics includes both artificial hops (longer arcs) and natural bounces with the surface (progressively shorter arcs). Velocity magnitude and direction are randomized to simulate the effect of local surface properties.}
	\label{fig:CG_bounces}
\end{figure}

Simulating the hopping motion of the probes around the small body calls for some considerations on the probe dynamics. A probe can leave the surface and enter a suborbital arc mainly because of two different events: either an active hop initiated by the probe itself, or a natural bouncing event caused by the probe's residual kinetic energy. In the latter case, the probe's bouncing velocity is largely unpredictable, as it is affected by local topography and surface absorption properties, which are usually heterogeneous throughout the small body. Figure \ref{fig:CG_bounces} shows the random nature of the hopping motion around a small body. As such, modeling the deterministic links between consecutive arcs proves challenging and would likely provide marginal gains for gravity estimation. In addition, the trajectory arcs produced by natural bounces may not be long enough to substantially enhance gravity knowledge. Hence, for gravity estimation, we only consider the arcs caused by the probe's active hopping events, i.e., we neglect all bounces of the probes with the surface. We assume that the arcs are uncorrelated with each other, and we process them separately for OD. 

\subsection{Target Body}

The population of small bodies in the solar system is highly diverse in many regards, such as composition, orbit, size, bulk density, and mass. Since the gravity signal is proportional to the small body's mass, this is a key parameter for gravity performance. Past missions have proven that higher-order gravity terms of small bodies are unobservable when gravimetry is performed by tracking an orbiting spacecraft. This is especially true for low-mass bodies. In order to consider the worst-case scenario for gravimetry, we assume that the simulated target body is a low-mass asteroid with the same mass, rotation rate, and pole as the asteroid Bennu visited by the OSIRIS-REx mission\cite{mazarico2017recovery}\cite{scheeres2019dynamic}. Table \ref{tab:small_body} captures the key parameters for the simulated small body, whereas its non-uniform gravity model is described in the dynamics section. We assume that the body orbits at 1 Astronomical Unit (AU) from the Sun.

\begin{table}[]
\noindent\scriptsize
    \begin{minipage}[t]{0.5\textwidth}
    \caption{Parameters of the simulated small body.}
    \begin{tabularx}{\textwidth}[t]{X X X}
    \toprule
         Parameter & Value & Units \\
         \hline
         $\mu$ & 4.892 & $\mathrm{m}^3/\mathrm{s}^2$\\
         $\mu_{\mathrm{mascon}}$ & 0.0087 & $\mathrm{m}^3/\mathrm{s}^2$\\
         Mascon [$\phi$,$\lambda$] & [45,45] & deg\\
         Radius & 246 & m\\
         Pole [RA,Dec,PM] & [85.65,-60.17,54.77] & deg\\
         Rotation Period & 4.2 & hours\\
         Albedo & 0.04 & --\\
         Emissivity & 0.9 & --\\
        \hline
    \end{tabularx}
    \label{tab:small_body}
    \end{minipage}%
    \begin{minipage}[t]{0.5\textwidth}
    \caption{Parameters of the notional probes.}
    \begin{tabularx}{\textwidth}[t]{X X X}
    \toprule
         Parameter & Value & Units \\
        \hline
        Shape & Sphere & --\\
        Radius & 3 & cm\\
        Mass & 0.2 & kg\\
        Albedo & 0 & --\\
        Emissivity & 0 & --\\
        Hop. $\Delta v$ Range & [3,14] & cm/s\\
        \hline
    \end{tabularx}
    \label{tab:probes}
    \end{minipage}
\end{table}

\section{Dynamical Environment}

The ground-truth probe trajectories are simulated accounting for the main forces acting in the small-body environment. We consider the central body's non-uniform gravity, N-body gravitational pulls, the solar radiation pressure (SRP), as well as the thermal radiation pressure (TRP) and the albedo pressure (AP) from the small body. Figure \ref{fig:probe_accelerations} shows the evolution of some representative accelerations as a function of the probe's altitude. In addition, stochastic accelerations are included to simulate the presence of unmodeled forces. This dynamics model is only used to simulate and estimate the probe's motion, as we assume that the observer, i.e. the mother spacecraft, is constantly hovering at a predefined altitude with no forces acting on it.

\subsection{Probe Design}

We consider a notional probe whose parameters are reported in Table \ref{tab:probes}. We consider the two probe designs proposed by Hockman et al. in the NIAC study: (1) an active, 5-cm, cubic-shaped probe with strobing LEDs to enable detection and (2) a passive, white-coated, spherical-shaped probe detected by passively reflecting the sunlight. The cubic design presents multiple advantages, such as reduced size (and hence reduced non-gravitational perturbations) and better photometric properties, thanks to the use of low-power LEDs which make the probe visible even when in the body's umbra. Hence, this second design is considered here for dynamics modeling. Since the probe is likely subject to fast rotation during its motion, we consider its interaction with radiation pressures as isotropic, and hence we approximate the cube with its circumscribing sphere. Furthermore, we do not account for reflected and infrared radiations from the probe, hence we set its albedo and  emissivity to zero. These assumptions are further discussed and justified in the radiation pressure section.  

\subsection{Gravity}

The gravity of major solar system bodies is considered. These are modeled as point-mass gravity sources using the \textit{de430} ephemeris data set\cite{folkner2014planetary}. 

The non-uniform gravity of the target small body is simulated using an artificial gravity field, since no estimates of higher degree and order for small bodies are available to date. We use an exterior spherical harmonic model, where the gravity potential can be written as the sum of the point-mass potential $\mu/r$ and a spherical harmonic expansion. In spherical coordinates $(r,\phi,\lambda)$ and with respect to a body-fixed reference frame, this can be written as shown in Equation \ref{eq:sph_harmonics}\cite{schutz2004statistical}.

\begin{equation}
\label{eq:sph_harmonics}
\begin{split}
U(r,\phi,\lambda) = & \frac{\mu}{r}+U'\\
U' = & -\frac{\mu'}{r}\sum_{l=1}^\infty \left(\frac{a_e}{r}\right)^lP_l(\mathrm{sin}(\phi))J_l+\\
& +\frac{\mu^*}{r}\sum_{l=1}^\infty\sum_{m=1}^l\left(\frac{a_e}{r}\right)^l P_{lm}(\mathrm{sin}(\phi))[C_{lm}\mathrm{cos}(m\lambda)+S_{lm}\mathrm{sin}(m\lambda)]
\end{split}
\end{equation}

where $\mu$ is the standard gravitational parameter of the central body, $r$ is the range from the central body's center of mass to the spacecraft of interest, $a_e$ and $\mu^*$ are the reference radius and the reference gravitational parameter used to normalize the harmonic coefficients $C_{lm}$ and $S_{lm}$. $P_{lm}$ are the Legendre Associated Functions of degree $l$ and order $m$. $J_l$, $C_{lm}$ and $S_{lm}$ are the normalized spherical harmonic coefficients, where $J_l=-C_{l0}$ represents the so-called zonal harmonics; $C_{lm}$ and $S_{lm}$ are referred to as sectoral harmonics when $l\neq m$ and as tesseral harmonics when $l=m$. The spherical harmonic expansion can be truncated to model the number of degree and orders of interest. 

\begin{figure}
\begin{minipage}{.49\textwidth}
  \centering
  \includegraphics[width=\linewidth]{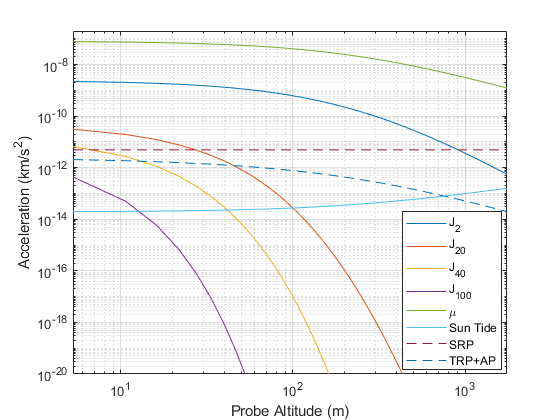}
  \captionof{figure}{Sample gravitational and non-gravitational accelerations as a function of altitude.}
  \label{fig:probe_accelerations}
\end{minipage}
\begin{minipage}{.49\textwidth}
  \centering
  \includegraphics[width=\linewidth]{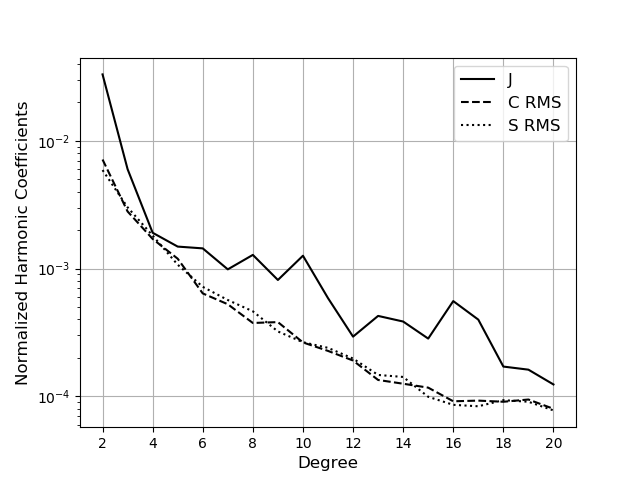}
  \captionof{figure}{Simulated harmonic coefficients of the ground-truth gravity field.}
  \label{fig:sh_truth}
\end{minipage}%
\end{figure}

We artificially generate the above harmonic coefficients. It has been shown that the amplitude of such parameters decays with degrees by following a power law, i.e., the so-called Kaula rule\cite{kaula2013theory}. McMahon et al. forward-modeled a modified Kaula rule for the asteroid Bennu, which is defined by Equation \ref{eq:kaula} and whose parameters are reported in Table \ref{tab:kaula}\cite{mcmahon2018osiris}. We generate the ground-truth gravity field by randomly sampling the harmonic coefficients from the Gaussian distribution associated with the Kaula rule for Bennu, up to degree and order 20. Figure \ref{fig:sh_truth} shows the resulting ground-truth gravity coefficients. As shown in Equation \ref{eq:sph_harmonics}, the number of coefficients for each degree increases with the degree itself, hence simulating a large number of trajectories using a gravity field larger than 20$\times$20 proved challenging and requires substantial computational resources. This will be addressed in future work.

\begin{equation}
\begin{split}
C_{l0}= & -J_l\approx\frac{K_{\mathrm{zonal}}}{n^{\alpha_{\mathrm{zonal}}}},\\ C_{lm}(m\geq 1) \;\mathrm{and}\; S_{lm}(m\geq 1)= & \sqrt{\frac{\sum_{m=1}^n(C_{lm}^2+S_{lm}^2)}{2l}}\approx\frac{K_{\mathrm{rms}}}{n^{\alpha_{\mathrm{rms}}}} 
\label{eq:kaula}
\end{split}
\end{equation}

\begin{table}[]
    \centering
    \caption{Mean and $1\sigma$ standard deviation of the Bennu's modified Kaula rule parameters.}
    \begin{tabular}{lll} 
           & K & $\alpha$ \\
        \hline
        Zonal  & $0.084\pm 0.021$ & $2.08\pm 0.13$ \\
        RMS  & $0.026\pm 0.005$  &  $2.01\pm 0.12$ \\
        \hline
    \end{tabular}
    
    \label{tab:kaula}
\end{table}

It is known that the exterior spherical harmonic expansion only converges when the position vector is located outside of the central body's Brillouin (circumscribing) sphere. Oftentimes, the shapes of small bodies present concave regions, hence said condition for orbital motion is not always satisfied. Several alternative gravity models have been proposed to model the spacecraft motion within the body's Brillouin sphere, e.g. the so-called interior models\cite{scheeres2016orbital}. To limit the scope of this study, we assume that the small body considered has a perfectly spherical shape. As such, The motion of the hopping probes around its surface can be fully represented with the exterior spherical harmonics model. Future work will include interior gravity models in the simulations.

In addition to the non-uniform gravity provided by the spherical harmonic model, we simulate the presence of a surface mass concentration (or \textit{mascon}) to assess the observability of gravity anomalies introduced by local surface feature, such as large boulders. We assume the mascon's mass to be a rough estimate of \textit{BenBen Saxum}'s mass (Table \ref{tab:small_body}), the largest boulder on the surface of Bennu.

\subsection{Radiation Pressure}

The probe acceleration given by the solar radiation pressure is computed as
\begin{equation}
    \ddot{\mathbf{r}}_{SRP}=\frac{f \Psi}{m r_\Sun^2}\mathbf{F}_N
\end{equation}
where $f$ is the shadowing scaling factor which is 1 when the probe is outside the shadow of the body, 0 when the probe is in the umbra of the body shadow, and between 0 and 1 when the probe is in the penumbra of the body shadow; $\Psi$ is the solar flux at 1 AU, in Newton; $m$ is the probe mass; $\mathbf{r}_\Sun$ is the position vector from the probe to the Sun []; $\mathbf{F}_N$ is the SRP geometry vector for a spherical body, computed as
\begin{equation}
    \mathbf{F}_N=-\pi R^2 \left(  1+\frac{4}{3}\kappa_d\nu_d \right)\hat{r}_\Sun
\end{equation}
where $R$ is the sphere radius, $\kappa_d$ is the diffuse reflectivity degradation factor and $\nu_d$ is the diffuse reflectivity factor. $\nu_d$ is defined as
\begin{equation}
    \nu_d=\frac{1}{3}(\gamma(1-\beta)+k(1-\gamma))
\end{equation}
where $\gamma$ is the fraction of incoming radiation that is reflected and $\beta$ is the fraction of reflected radiation that is specular. Since we assume $\gamma=0$ (i.e., the probe albedo is zero), we only consider the effect of direct radiation pressure. Hence, $\ddot{\mathbf{r}}_{SRP}$ can be rewritten as
\begin{equation}
    \ddot{\mathbf{r}}_{SRP}=f \Psi\frac{A}{m}\frac{\hat{r}}{r_\Sun^2}
\end{equation}
which is often called the cannonball model.

We use a standard cannonball model to account for SRP accelerations, which assumes the body is a sphere\cite{mcmahon2020dynamical}. This is usually a reasonable assumption even for non-spherical bodies subject to fast tumbling, such as the notional cubic probes, since the anisotropic SRP contributions even out over time. We assume that the body has zero albedo and emissivity, i.e. the effect of the SRP only accounts for the direct photon flux, and not for reflections nor re-emitted infrared radiation. Note that this assumption mostly affects the ground-truth trajectory and not gravity estimation results.

The small body is also a source of radiation pressure, with a component given by the reflected sunlight (the albedo pressure, or AP), and one from the thermal re-emission due to the body's heating from the Sun (the thermal radiation pressure, or TRP). In this case, the AP is given by:
\begin{equation}
\begin{aligned}
    \ddot{\mathbf{r}}_{AP} = \iint_S\left(\frac{\Psi}{r_\Sun^2} \frac{A}{m} a(\mathbf{r}_S)\,\mathrm{cos}(\Phi_S)\frac{1}{\pi r_S^2}\hat{r}_\Sun \right) dS
\end{aligned}
\end{equation}
where $S$ is the surface of the spherical cap visible by the spacecraft, $a(\mathbf{r}_S)$ is the surface albedo corresponding to the surface point $\mathbf{r}_S$ and $\Phi_S$ is the angle that the sunlight direction makes with the surface normal at $\mathbf{r}_S$. The same expression can be used to compute the albedo pressure component given by the small body's thermal emission, by substituting the albedo $a(\mathbf{r}_S)$ with the emissivity function $e(\mathbf{r}_S)$. Here, we assume that both the albedo and the emissivity are constant parameters throughout the body surface (Table \ref{tab:small_body}).

\subsection{Stochastic Accelerations}

In addition to the deterministic forces described above, stochastic accelerations are included in the dynamics model to compensate for both unmodeled forces and modeling errors in the dynamics. As discussed by Chesley et al. for the OSIRIS-REx particle tracking case, there is a number of other relevant forces affecting the motion in the vicinity of the small body, such as the SRP and AP radiation reflected and re-emitted by the probes and the Poynting-Robertson effect, due to stellar aberration affecting the incoming SRP\cite{chesley2020trajectory}. In this case study, mismodeling of the probe dynamics may be caused by residual probe outgassing as well as changes in probe reflectance and emissivity properties, due to repeated interactions with the surface regolith.

We model stochastic accelerations as an uncorrelated white random process with a given mean and standard deviation, divided into consecutive batches. For estimation, each batch can have a different acceleration mean, i.e.:
\begin{equation}
    \hat{\mathbf{a}}=\{\hat{\mathbf{a}}_1,...,\hat{\mathbf{a}}_n\}\;\; \mathrm{for}\;\; \Delta t = \{ \Delta t_1, ..., \Delta t_n\}
\end{equation}
where $\hat{\mathrm{a}}$ are the estimated accelerations and $\Delta t$ are the corresponding $n$ batch intervals. In the ground-truth dynamics model, we assume that all batches are zero-mean. In the OSIRIS-REx scenario, the unmodeled forces were always smaller than $1\times10^{-11}\mathrm{km}/\mathrm{s}^2$\cite{chesley2020trajectory}. As such, we set the stochastic accelerations' standard deviation as $\sigma_{\mathrm{st}}=1\times10^{-11}\mathrm{km}/\mathrm{s}^2$ to make a conservative assumption on the unmodeled forces. Such stochastic accelerations also account for errors in modeling the SRP and AP, whose magnitude is always below said value (Figure \ref{fig:probe_accelerations}). The batch interval is set to 10 minutes. Since the batch interval is a tunable parameter for estimation, future work will focus on more rigorous tuning strategies to optimize OD performance.

\subsection{Hopping Events}

As previously discussed, all arcs considered in this study are initiated by a hopping event: the probe, initially steady on the terrain, activates the hopping mechanism which manifests as a $\Delta v$ with respect to the small body surface. Even if we assume that the hopping mechanism is calibrated to provide a constant impulse, both the magnitude and direction of the hopping velocity is randomized by the heterogeneous properties of the small body surface. The hop velocity magnitude depends on the local absorption properties of the surface regolith, which can vary from dusty to rocky regions. The velocity direction is largely affected by the terrain topography, which can feature slopes and obstacles and also varies across the surface. 

To model these effects in our simulations, we randomly sample the hopping velocity as follows. The velocity magnitude is sampled from a uniform distribution:

\begin{equation}
    v_{\mathrm{hop}}=\mathcal{U}(3,14)\;\mathrm{cm/s}
\end{equation}

The distribution range is selected via trial and error to enable arcs with good gravity observability (especially for the higher-degree and order terms). The direction of the hopping velocity is defined using its right ascension (RA) and declination (Dec) Euler angle, defined with respect to a local-vertical local-horizontal frame. These angles are also sampled using uniform distributions:

\begin{equation}
\begin{cases}
RA_\mathrm{hop} = 2\pi u, \; \mathrm{where} \; u=\mathcal{U}(0,1)\\
Dec_\mathrm{hop}=\mathrm{cos}^{-1}(\nu), \; \mathrm{where} \; \nu=\mathcal{U}(0,0.5)\\
\end{cases}
\end{equation}
where the declination angle can span between 0 and 60 degrees. The $\mathrm{cos}^{-1}$ formulation is used to avoid biasing the velocity. (If spherical coordinates are directly sampled from a uniform distribution, the resulting direction is effectively biased toward the pole(s).) 

The initial position of the probes is also randomized to ensure that the trajectory arcs spread across the entire body. The latitude $\phi_0$ and longitude $\lambda_0$ of the probe's initial position are randomly sampled using the same approach previously described, whereas the radial coordinates is set equal to the small body radius.

\subsection{Arc Population}

\begin{figure}[]
	\centering\includegraphics[width=.5\textwidth]{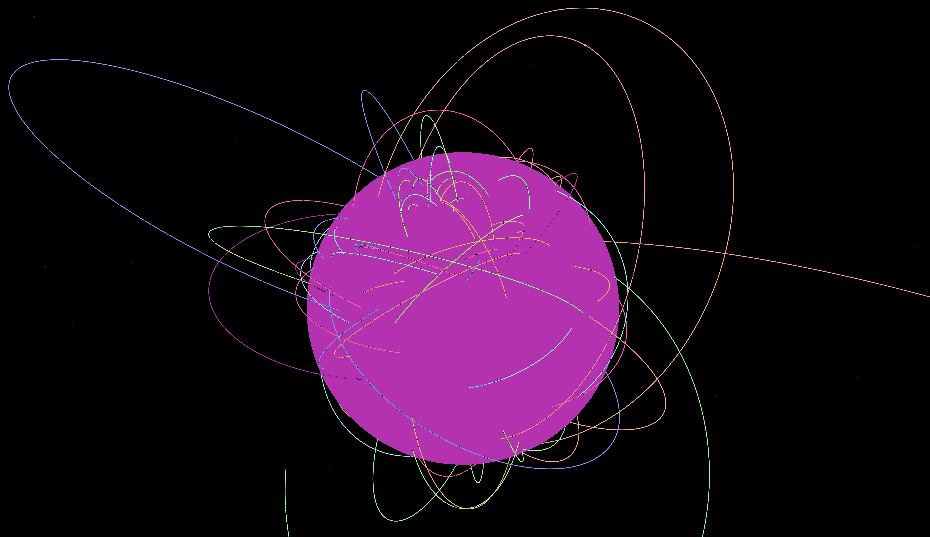}
	\caption{3D visualization of 100 sampled probe arcs (colored trajectories) around the small body (purple sphere) in the asteroid-centered inertial frame.}
	\label{fig:traj_3D}
\end{figure}

\begin{figure}[]
	\centering\includegraphics[width=\textwidth]{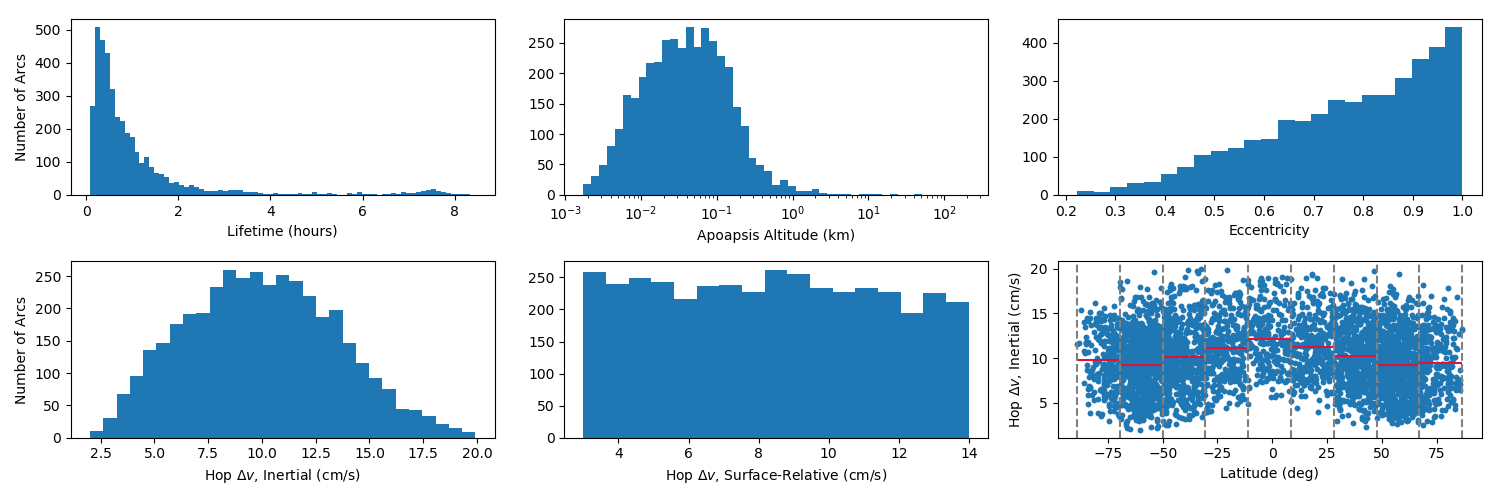}
	\caption{Histograms for the simulated arc parameters. Hyperbolic trajectories are not reported. The red bar indicated the mean value for each bin.}
	\label{fig:arc_hist}
\end{figure}

Using the above considerations, we simulate a baseline data set with 4000 randomized trajectory arcs around the target body. A sample of 100 arcs is visualized in Figure \ref{fig:traj_3D}, whereas some descriptive parameters of the resulting data set are shown in Figure \ref{fig:arc_hist}. The mean arc lifetime is about 1 hour and the altitude of the apoapsis with respect to the surface is between tens and hundreds of meters, for most cases. The idea behind such short, low-altitude arcs is to maximize the time spent close to the surface, where the gravity signal is stronger (Figure \ref{fig:probe_accelerations}). Additionally, we note that most arcs feature high eccentricities. This is arguably because suborbital arcs always present subsurface periapsis, as discussed by McMahon et al.\cite{mcmahon2020dynamical}

The small body rotation affects the probe arcs by changing the probe's velocity in the inertial frame. This effect is more relevant for fast-rotating bodies and when the probes are located at low latitudes, as shown in Figure \ref{fig:arc_hist}. Furthermore, the hopping velocity is always biased toward the east direction. This effect could be mitigated by actively controlling the hopping velocity according to the latitude the probe is located at.

A very small fraction of the arc population can result in escape trajectories. For this data set, about 0.3\% of the orbit becomes hyperbolic and escape from the small body. While the hopping velocity can be easily designed to avoid direct escape from the body, it has been shown that escape events can also occur after periapsis passage due to the perturbed dynamical environment \cite{mcmahon2020dynamical}. This is more likely for higher-altitude arcs, where perturbations act for a longer time. In this study, arcs evolving into hyperbolic orbits are not considered. Future work will assess in detail escape statistics and related design choices.

\section{Probe Measurements}

Gravity estimation is performed by tracking the probe trajectories from an observing mother spacecraft, using an onboard camera that provides relative-angle measurements. These are the only data processed for OD, i.e., no ground-based radiometric measurements are considered.

\subsection{Mother Spacecraft}
We assume that the mother spacecraft is hovering relative to the small body, at a 4 km distance from the small body center and at a zero Sun phase angle (i.e., the Sun is behind the observer). The altitude is chosen such that the small body spans approximately 80\% of the camera field-of-view width, which is considered a good trade-off between probe resolution and width of the observed scene. 

The spacecraft position is modeled as a constant, inertial offset with respect to the small body center. To compensate for the lack of the observer's dynamics model (e.g., uncertainties introduced by station-keeping and other perturbations), we implement a position's consider covariance which is processed by the OD filter (as discussed in the filter setup section). Future work will also tackle trajectory design for the mother spacecraft. The spacecraft attitude is nadir-pointing, with a white-noise stochastic pointing error associated to a 20-minute batch interval. The stochastic model is the same as the one used for stochastic accelerations. 

Under the assumption of a low-budget mission, we define the camera model based on the JPL's ASTERIA camera, a small yet capable instrument used to perform astrophysical measurements from aboard a CubeSat\cite{pong2019camera}. The parameters used for the camera model are reported in Table \ref{tab:asteria}. The medium field of view enables global coverage of the small body surface while observing from a safety distance, whereas the high resolution ensures high-quality astrometry. We assume a pinhole camera model, neglecting the effect of distortion and other error sources\cite{owen2011methods}, though periodic camera calibration can be used to model distortions in practice.

\begin{table}[h!]
    \centering
    \caption{Camera model parameters.}
    \begin{tabular}{ccc}
        Parameter & Value & Units \\
        \hline
        Aperture diameter & 60.7 & mm\\
        Focal length & 85 & mm\\
        Detector dimensions & $2592\times 2192$ & pixels\\
        Pixel size & $6.5\times 6.5$ & $\mu$m\\
        Detector Field of View & $11.2\times 9.6$ & deg\\ 
        Measurement Uncertainty & 0.1 & pixels\\
        \hline
    \end{tabular}
    \label{tab:asteria}
\end{table}

\subsection{Probe Visibility}

One strength of using bright artificial probes is that they can be detected not only on the dark background, but also when moving in front of the small body surface. In this second case, the probes altitude is usually lower (e.g. see Figure \ref{fig:traj_3D}) and hence the gravity signal stronger. To assess the quality of probe detections in front of the surface we consider its photometric and astrometric properties. The probe's photometry is given by comparing the relative magnitude of the photon flux emitted by the probe with the one from the small body surface. The astrometric uncertainty represents the detection error given by the spatial distribution and discretization of the probe signal in the camera plane. Hockman et al. presented a probe visibility analysis for the \textit{Gravity Poppers} scenario, considering the same mission scenario presented in this work, and concluded that both passive probes and strobing probes can be detected in front of the small body surface with an photometric SNR of 2 or more. The astrometric error presented in said work is on the order of hundredth of a pixel; however, it does not account for camera calibration and other error sources. Hence, we conservatively assume an astrometric error of 0.1 pixel, which matches the measurement uncertainty in the OD filter (Table \ref{tab:asteria}). Future work will further analyze how photometric and astrometric errors affect the measurement uncertainty for probe detections.

\subsection{Tracking Frequency}

In addition to measurement precision, tracking frequency also plays an instrumental role in gravity observability. As shown in Equation \ref{eq:sph_harmonics}, both the amplitude and the ``wavelength" of the gravity harmonics decrease with increasing degree. As such, observing higher-degree gravity harmonics requires (1) taking measurements with high spatial resolution with respect to the surface and (2) observing the motion at a relatively low altitude. 

State-of-the-practice techniques for optical navigation require all navigation images to be downloaded to the ground, in order to manually process surface landmarks. Given the limitations in the spacecraft-to-ground link data rate, there is an upper limit to the achievable imaging frequency, usually in the order of one image every 10 minutes. As discussed in the results section, such a tracking frequency greatly limits the amount of observable gravity terms, even when using low-altitude detections. Since the probe photometric properties make them easy to detect in front of the small body surface, we claim that an autonomous detection algorithm (e.g. based on thresholding or match filtering) could be implemented onboard. This routine could extract the probe centroids and download them to the ground, in place of the entire image, for OD and gravity estimation purposes. This process would dramatically reduce the amount of data volume required, and hence would enable much higher operational tracking frequencies. 

\begin{figure}[htb]
	\centering\includegraphics[width=0.7\textwidth]{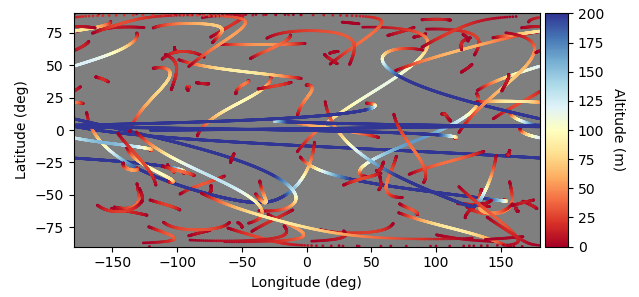}
	\centering\includegraphics[width=0.7\textwidth]{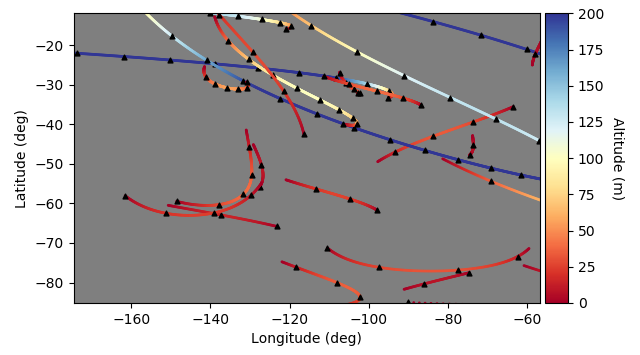}
	\caption{Scatter plots of the observed probe positions in the small body-fixed frame. Top: the entire latitude-longitude domain. Bottom: a magnified image comparing the 10-minute tracking interval (black triangles) with the 10-second interval (colored dots). The altitude color bar is saturated at 200 meters.}
	\label{fig:traj_groundtrack}
\end{figure}

With this reasoning, we assume that our notional spacecraft can sample one image every 10 seconds. This parameter will also be refined in future work by including more system-level considerations. For the arc population considered in this scenario, the difference in spatial resolution between a 10-minute and a 10-second sampling interval is massive, as shown by the ground tracks in Figure \ref{fig:traj_groundtrack}. Importantly, the advantage of increasing the tracking frequency is twofold: a higher spatial coverage of the gravitational potential and the increased number of measurements for each arc. Since the apoapsis altitude of most trajectory is very low, a 10-minute imaging interval would only provide a handful of measurements for each arc, which limits the observability of high-degree gravity terms. 

\section{Covariance Analysis}

In order to quantify gravity estimation performance, we perform a covariance analysis on the probe tracking scenario. To do so, we use the ground-truth dynamics to simulate perfect measurements of the probes. Each estimated parameter is associated with an a-priori covariance, and measurement uncertainty is accounted. 

\subsection{Multi-Arc Filter}

We use a \textit{multi-arc} filtering approach to treat the uncorrelated trajectories spanning over multiple time intervals\cite{french2020precise}. This technique has been effectively used in flight operations, e.g. to perform particles OD at the asteroid Bennu and to estimate the Earth and Moon gravity field with unprecedented precision, during the GRACE and GRAIL missions, respectively\cite{french2020multi,asmar2013scientific}. 

The key principle of the multi-arc filter is that the estimated quantities can be divided into \textit{local} and \textit{global} parameters, where the former are valid only for a specific OD fit and the latter affect all OD solutions. In this study, the probe states for each arc can be considered the local parameters, whereas all the small body-related parameters (gravity field, pole, etc.) are global. By framing the problem in this way, each arc's OD can be performed separately. Then, each OD solution can be used to recursively update the global parameters estimates and related uncertainties. The estimates of the global parameters improve and the uncertainties decrease as more arcs are processed. For a covariance analysis, we are only interested in the parameters uncertainties and not the estimated values since the measurement residuals are always zero. In particular, we are interested in the covariance evolution of the gravity field coefficients. The gravity covariance obtained by processing multiple arcs is computed using the following fundamental property of information matrices: if $N$ different information matrices are produced using $N$ different sets of observations, and the observation sets are independent, then the total information matrix can be written as the sum of each information matrix. Then, the estimated gravity covariance matrix $\hat{\mathbf{P}}_G$ can be written as

\begin{equation}
\label{eq:multi_arc}
    \hat{\mathbf{P}}_G=\left(\mathbf{P}_{G,0}^{-1}+\sum_{j=1}^N\Lambda_{G}^j\right)^{-1}
\end{equation}

where $\Lambda_G^j$ is the gravity information matrix given by the $j$-th set of observations (i.e. by the $j$-th arc) and $\mathbf{P}_{G,0}$ is the a-priori gravity covariance matrix. Note that this expression is only valid if the arcs are uncorrelated with each other. Moreover, it is important to avoid that the a-priori information is counted for every single arc, as this would bias the information gain toward higher, unrealistic values. 
The multi-arc filtering process can be summarized as follows. After defining the a-priori covariance for the global parameters, for every arc:
\begin{enumerate}
    \item random initial position and velocity, together with the a-priori arc covariance, is defined (as described in the hopping event section);
    \item the ground-truth trajectory and related partials are propagated;
    \item perfect measurements are simulated, based on the ground-truth trajectory;
    \item the posterior covariance for the local and global parameters is computed;
    \item the gravity-field covariance is updated based on the information from the given arc.
\end{enumerate}

\subsection{Filter Setup}
The OD process is performed using a Square Root Information Filter (SRIF)\cite{schutz2004statistical}. Table \ref{tab:filter_params} shows the parameters processed by the filter and related a-priori uncertainties, whose values are assumed very large, compared to the nominal value, to avoid double-counting the information in the multi-arc filter, as previously discussed. In this way, the information gain can be considered data-driven. The probe trajectories are dynamic parameters, whereas all the small body-related quantities are assumed to be constant values (i.e. bias parameters). The position uncertainty of the observing spacecraft is modeled as a 5-meter consider covariance, which is in line with position uncertainty values obtained during landmark-based navigation in proximity of the small body, for flight scenarios\cite{leonard2019osiris}. The stochastic pointing uncertainty of 0.003\,deg is in line with the OSIRIS-REx case.
Finally, we assume that some subset of the probes that are at rest on the surface can be used as body-fixed landmarks. Due to their relative brightness with respect to the surface, steady probes can provide high-quality surface-relative measurements to enhance the gravity estimation. Here, we assume that 10 probes are used as landmarks for any given arc and we randomly sample the latitude and longitude of such landmarks in the equatorial band corresponding to $\phi\in[-20^o,+20^o]$.

\begin{table}[]
    \caption{A-priori uncertainties for estimated parameters}
    \centering
    \begin{tabular}{c|c|c|c}
        Name & Type & A-Priori $\sigma$ & Units\\
        \hline
        Probes pos. & Dynamic & 100 & km \\
        Probes vel. & Dynamic & 100 & m/s \\
        Pole RA & Bias & 100 & deg\\
        Pole Dec & Bias & 100 & deg\\
        Pole PM & Bias & 100 & deg\\
        Pole $\omega$ & Bias & 100 & deg/day\\
        $\mu$ & Bias & 1 & $\mathrm{km}^3/\mathrm{s}^2$\\
        $\mu_{\mathrm{mascon}}$ & Bias & 1 & $\mathrm{m}^3/\mathrm{s}^2$\\
        Normalized SH coeff. & Bias & 1 & --\\
        Landmarks & Bias & 100 & m\\
        S/C pos. & Consider & 5 & m\\
        Stoch. Accel. & Stochastic & $\mathrm{1\cdot 10^{-11}}$ & $\mathrm{km/s^2}$\\
        Stoch. Pointing & Stochastic & 0.003 & deg\\ 
        \hline
        
    \end{tabular}
    \label{tab:filter_params}
\end{table}

\section{Gravity Estimation Results}

In this section, we present gravity estimation performance for the above simulated scenario. The quality of gravity estimates depends on several design parameters and on the duration of the observation campaign. We firstly present gravimetry results obtained by processing all the 4000 simulated arcs. Second, we discuss the evolution of gravity information over time and use a back-of-the-envelope analysis to infer required mission durations. Third, we assess the sensitivity of performance to key mission and arc parameters.

\subsection{Gravity Field Observability}

\begin{figure}[htb]
	\centering\includegraphics[width=0.49\textwidth]{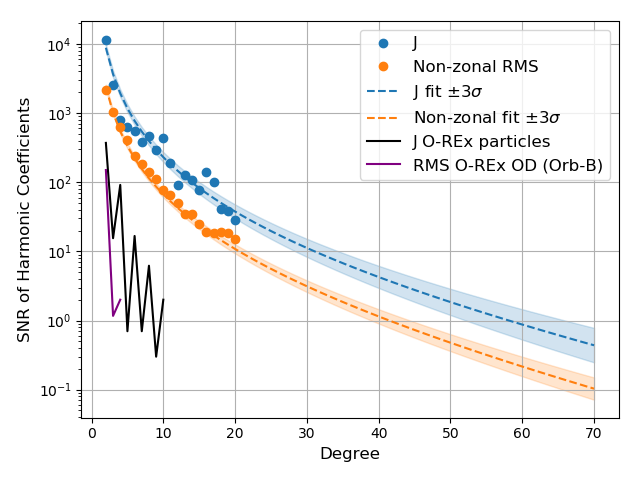}
	\centering\includegraphics[width=0.49\textwidth]{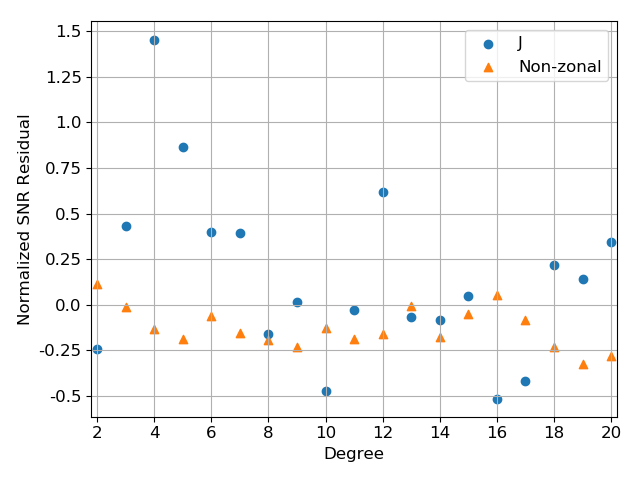}
	\caption{SNR of the estimated harmonic coefficients. Left: comparison between the coefficients SNR estimated in this work (blue and orange) and by the OSIRIS-REx mission operations using particle tracking (black) and spacecraft tracking (purple). The dots are the SNR obtained by OD and the dashed lines represent the estimated model for SNR evolution. Right: residuals between the estimated model for SNR evolution and the actual SNR obtained by OD.}
	\label{fig:sh_spectrum}
\end{figure}

Gravimetry results obtained by processing all the simulated arcs are shown in Figure \ref{fig:sh_spectrum} (left). The metric we use to assess performance is the signal-to-noise ratio (SNR) of the estimated harmonic coefficients, which is computed as the coefficient's ground-truth magnitude over the posterior 1-$\sigma$ for that same coefficient, provided by the OD solution. Note that the posterior uncertainty is given by the sum of all information matrices from each processed arc, as shown in Equation \ref{eq:multi_arc}. We separate the SNR of the zonal coefficients from the remaining (tesseral and sectoral) coefficients, as the former are systematically larger (which can be deduced by the parameters in Table \ref{tab:kaula}). The tesseral and sectoral terms (here defined \textit{non-zonal}) are considered as a whole using their RMS value.

We compare our gravimetry results with those obtained during the OSIRIS-REx mission, from both the particle tracking campaign and the standard spacecraft-tracking OD during the Orbital-B mission phase\cite{leonard2019osiris}. Results suggest that, using the proposed approach, the SNR for the zonal harmonic terms increases by about two orders of magnitude with respect to the OSIRIS-REx's particle tracking campaign, which could estimate the gravity field up to about degree 10, and by three orders of magnitude when compared to OSIRIS-REx's spacecraft tracking, which only estimated the $J_2$ term. It is important to note that these three cases differ in a number of ways, such as the amount and type of processed measurements, hence the comparison can only be made qualitatively. Finally, the mascon's SNR is 107, proving that even surface features such as large boulders, and potentially subsurface voids, can be observed with low uncertainties. 

Since the SNR of all simulated coefficients is always above the observability threshold (i.e., SNR$>$1) for all 20 degrees and orders, we fit a model to our SNR data to assess observability for higher-degree and order coefficients. The numerator of the SNR model is given by the Kaula's harmonic magnitudes (Equation \ref{eq:kaula}). For the denominator, we fit an exponential function $\alpha e^{\beta x}$ to the posterior 1-$\sigma$ data. The estimated parameters are reported in Table \ref{tab:exp}. This regression approach suggests that, under these assumptions, the zonal harmonic coefficients are observable up to about degree 60, whereas the remaining coefficients are observable up to degree 40, approximately. The residuals between the fitted model and the actual SNR obtained via OD is shown in Figure \ref{fig:sh_spectrum} (right). The residual values are normalized with respect to each degree's \textit{estimated} SNR (i.e. using the harmonic value estimated by OD) and have mean of 0.16 and -0.13 for zonal and non-zonal terms, respectively. The model appear to be valid to infer crude estimates of the highest-degree observable harmonics; however, future work will validate and refine this model by simulating higher-degree and order gravity fields.

\begin{table}[]
    \centering
    \caption{Estimated parameters of the exponential model $\alpha e^{\beta x}$ for the posterior 1-$\sigma$ evolution.}
    \begin{tabular}{lll} 
           & $\alpha$ & $\beta$ \\
        \hline
        Zonal & $2.0975\times10^{-6}$ & $3.6937\times10^{-2}$\\
        RMS & $2.4817\times10^{-6}$ & $4.2667\times10^{-2}$\\
        \hline
    \end{tabular}
    
    \label{tab:exp}
\end{table}

\subsection{Mission Duration}
To implement the proposed gravity estimation concept in an actual mission, assessing requirements on the mission duration is paramount. Usually, proximity operations around a small body require continuous ground-based support, navigation, and planning, hence the probe observation time frame is strictly linked to the overall mission cost and complexity. Since our OD pipeline evaluates the probe arcs separately, the underlying mission duration is not explicitly defined. However, we can use the arc population statistics and a back-of-the-envelope approach to roughly quantify the mission duration. 

\begin{figure}[htb]
	\centering\includegraphics[width=0.49\textwidth]{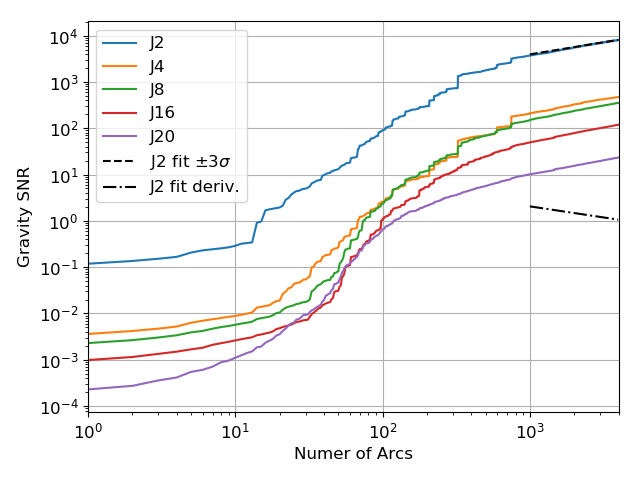}
	\centering\includegraphics[width=0.49\textwidth]{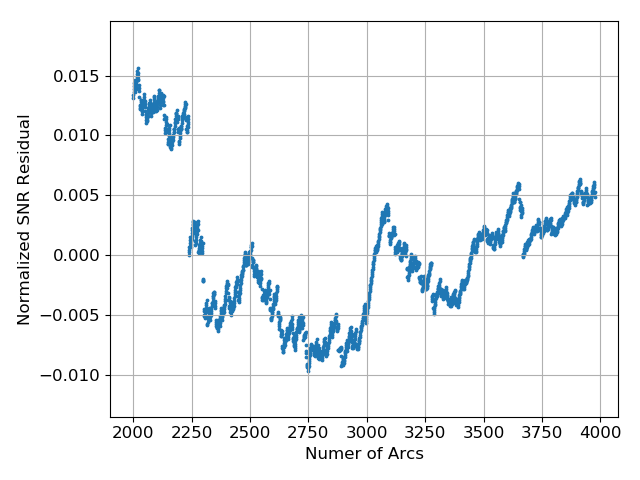}
	\caption{SNR evolution for some representative gravity harmonics. Left: actual evolution compared to the fitted model. Right: residuals of the actual values compared to the model.}
	\label{fig:sh_evol}
\end{figure}

We start by analyzing the gravity information gain as a function of the number of arcs. We fit the SNR evolution with a power law model $\alpha x^\beta$, as shown in Figure \ref{fig:sh_evol} (left). For example, the estimated power-law parameters for $J_2$ are $\alpha=64.4709$ and $\beta=0.5590$. Note that there is a transient state in SNR in the first hundreds of arcs, which is likely due to the fact that the surface coverage with few arcs is sparse and the information gain for each arc is more dependent upon the location of the arc itself. As such, we fit the SNR evolution starting from the 2000-th arc onward. The model residuals are lower than 2\% of the SNR value (Figure \ref{fig:sh_evol}, right). However, future work will simulate more arcs to validate this model.

\begin{figure}[htb]
	\centering\includegraphics[width=0.5\textwidth]{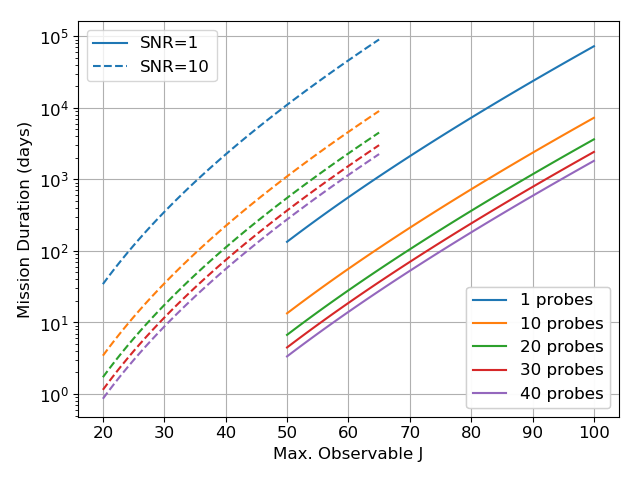}
	\caption{Mission duration as a function of the maximum observable zonal harmonic coefficient J. The plot is parametrized with the number of hopping probes.}
	\label{fig:mission_time}
\end{figure}

The power-law model and its derivative, shown in Figure \ref{fig:sh_evol} (left) suggest that there is a diminishing return in the information gain when the arc number increases. We use this relationship to assess the required mission duration as a function of the maximum observable zonal harmonic term, as shown in Figure \ref{fig:mission_time}. In other words, we assume a given gravity science objective (e.g. observing the zonal harmonics up to degree 40) and estimate a mission duration needed to achieve the goal. The mission duration $T_\mathrm{mission}$ is evaluated as:

\begin{equation}
    T_\mathrm{mission}=\frac{\bar{T}_\mathrm{arc}N_\mathrm{arc}}{f_\mathrm{vis}n_\mathrm{probe}}
\end{equation}

where $\bar{T}_{\mathrm{arc}}=69$ minutes is the mean arc lifetime (Figure \ref{fig:arc_hist}), $f_{\mathrm{vis}}=0.5$ is a visibility factor assuming that only half of the arcs are visible from the mother spacecraft, at any given time (due to the body occlusion; note that this is a conservative assumption); $n_{\mathrm{probe}}$ indicates the number of hopping probes utilized during the mission (which does not account for probes potentially used as steady surface landmarks). Results show that good estimates (i.e. SNR$>$10) up to $J_{40}$ can be obtained within a period spanning from a few months to about one year, depending on the number of probes used. On the other hand, the mission time increases as a power law with the target degree to observe, hence estimating higher-degree terms may not be done within reasonable time frames. In addition, there is little advantage in increasing the number of hopping probes above 20-30.

\subsection{Sensitivity To Measurements}

The presented mission scenario is based on a number of assumptions on the observation type and performance. In this section, we assess gravimetry performance for cases where such assumptions are relaxed or new types of observations are introduced. We simulate 1000 arcs for 5 different case studies: (1) the baseline simulation setup (described in the previous sections), (2) increasing the imaging rate to one image every 10 minutes, (3) increasing the optical measurement uncertainty to 0.5 pixels, (4) including ranging and Doppler measurements in the OD process, and (5) increasing the camera pointing uncertainty. Case 2 is used to simulate performance when the probes (or particles) are detected on the ground, and hence the imaging rate is limited by the downlink data volume (e.g. OSIRIS-REx particle tracking). Case 3 assumes that the photometric properties of the observed targets are poorer than the artificial probes', e.g. when tracking natural particles are tracked. In case 4, we treat the artificial probes as Two-Way coherent transponders and assume that a Deep-Space Atomic Clock (DSAC) is used aboard the mother spacecraft\cite{ely2018using}. We assume $\sigma_{\rho}=10\,\mathrm{cm}$ and $\sigma_{\dot{\rho}}=0.1\,\mathrm{mm/s}$ for the range and range-rate measurement uncertainty, respectively. We model such measurements as perfect range and range rate. Note that, for this measurement strategy to be adopted, a more advanced probe design is required to enable the spacecraft-to-probe link.

\begin{figure}[htb]
	\centering\includegraphics[width=0.5\textwidth]{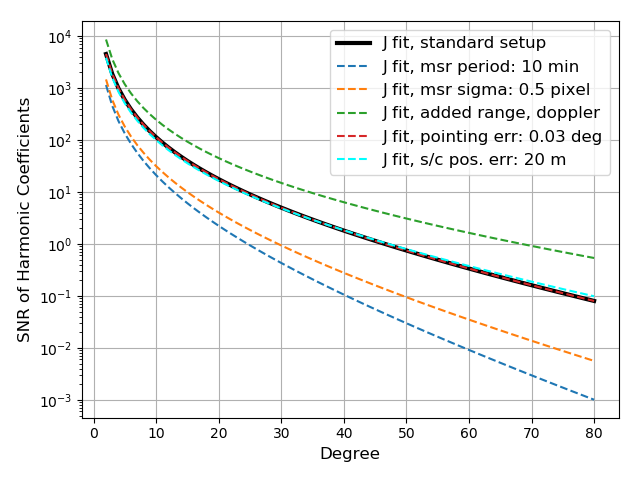}
	\caption{Fitted SNR model for the estimated zonal harmonics, when different observation parameters and when adding range and Doppler measurements.}
	\label{fig:sh_spectrum_sens}
\end{figure}

For each case, we fit the harmonic SNR with the exponential model previously introduced; results are shown in Figure \ref{fig:sh_spectrum_sens}. While all scenarios accurately recover the low-degree harmonics, the maximum observable degree varies widely among cases. As expected, introducing range and range-rate measurements outperforms all other scenarios, observing the zonal harmonics up to about degree 70, compared to degree 50 from the baseline. On the other hand, reducing the imaging frequency and measurement uncertainty decreases the maximum observed harmonic below 25 and 30, respectively. These conclusions highlight the importance of autonomous onboard probe detection and good photometric properties of the targets. Finally, gravimetry performance is not particularly sensitive to the magnitude of the stochastic pointing errors, and remain almost unchanged when the error is increased by an order of magnitude. Future work will include more thorough sensitivity analyses that explore a broader domain of the trade space for mission design.

\subsection{Sensitivity to Probe Trajectories}

\begin{figure}[htb]
	\centering
	\includegraphics[width=\textwidth]{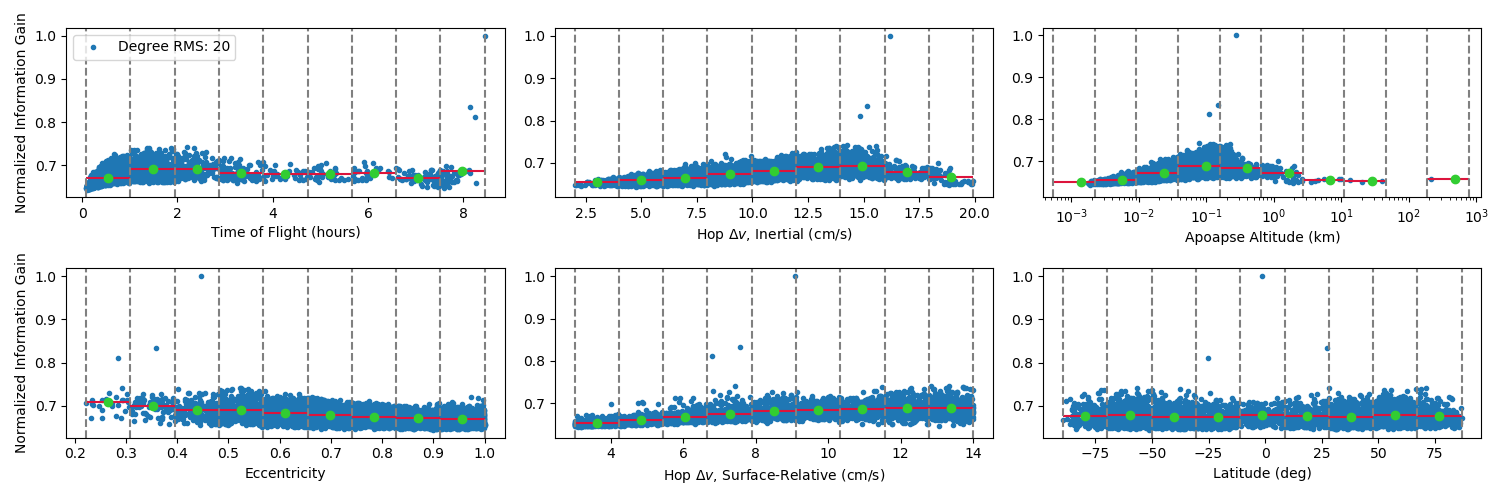}
	\includegraphics[width=\textwidth]{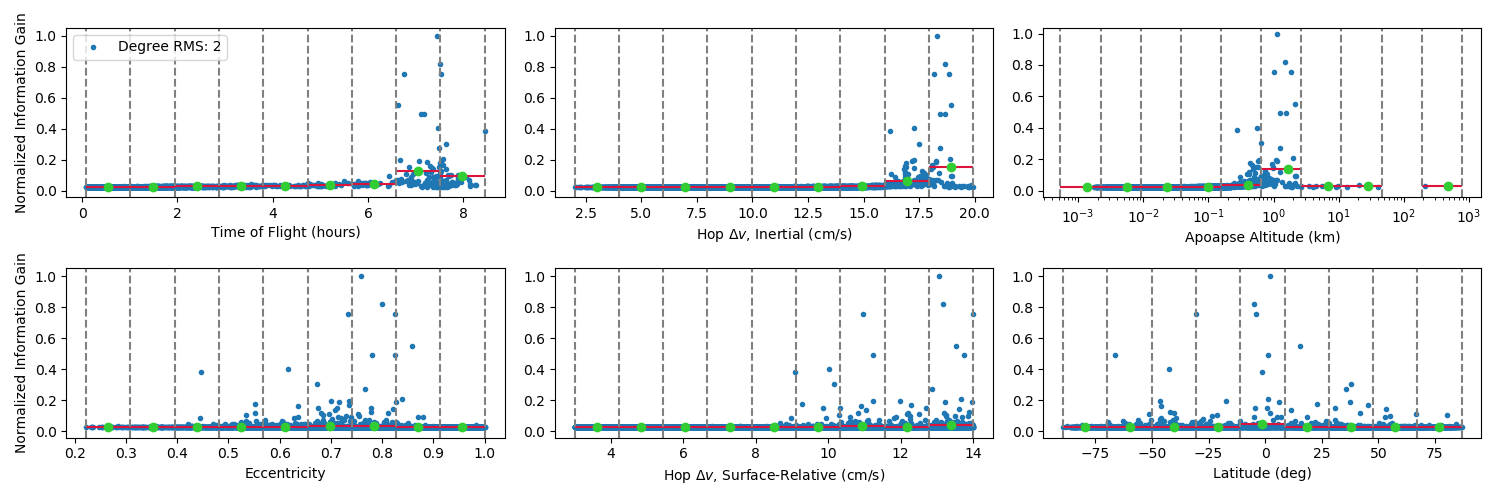}
	\caption{Gravity information gain as a function of selected orbital parameters, for degree 20 (top) and degree 2 (bottom). The red bar indicates the mean for the given data bin.}
	\label{fig:traj_sens}
\end{figure}

Finally, we analyze the sensitivity of gravimetry performance to some representative arc's orbital parameters. We selected the arcs lifetime, hopping velocity (both in the inertial and body-fixed frame), the altitude of the orbital apoapsis, the eccentricity, and the initial latitude. The information gain is represented for a given arc and a given harmonic degree, and is computed as the RMS of all gravity SNR for that degree (accounting for zonal and non-zonal coefficients with such a degree). Figure \ref{fig:traj_sens} presents results for degrees 2 and 20. As expected, gravimetry for the lower degree (which has a lower spatial resolution) is better for higher altitudes and hopping velocities, whereas the opposite is true for higher-degree terms. We also observe that the information gain variation for low-degree harmonics is much larger for low-degree terms, compared to higher degrees. This might be due to higher-altitude orbits being more diverse, as the orbital perturbations act on a longer time of flight. However, excluding some outliers, orbital parameters have little impact on gravity information gain for the vast majority of the arc population. The hopping velocity can be calibrated, hence it is arguably the most interesting parameter from a mission design standpoint. However, there is little advantage in fine-tuning this parameter to a specific value as the correlation with gravimetry performance is weak, especially for the (most interesting) high-degree gravity terms.

\section{Conclusions}

In this study, we assess gravity estimation performance at small bodies for a novel mission architecture. We reproduced a scenario similar to the OSIRIS-REx mission, around a low-mass small body such as asteroid Bennu. We used conservative assumptions on unmodeled forces and measurement uncertainty, but assumed that autonomous onboard probe detection is available to enable high tracking frequency. Results suggest that it is possible to recover the gravity field of the low-mass body up to degree 40 within days or months, depending on the target SNR, using 20 hopping probes. The mission time frame largely depends on the gravity science objective. For the asteroid Bennu, the gravity SNR outperforms spacecraft-tracking techniques by three orders of magnitude. The gravity contribution from surface features, such as large boulders, can also be precisely estimated. High measurement frequency and good probe photometric properties are key to achieve high gravimetry performance within a limited time frame, and spacecraft-to-probe radiometric measurements substantially enhanced the gravity estimates. On the other hand, gravimetry is not particularly sensitive to the orbital parameters of the hopping probes.

Future work will increase simulation realism, e.g. including the mother spacecraft orbital dynamics, using non-spherical models for the target body shape, and interior gravity models. Additionally, we will extend sensitivity analyses to a broader trade space of mission design and measurement strategies. Finally, we will implement a density estimation scenario to assess how this gravimetry technique could be applied for specific planetary science objectives.

\section{Acknowledgements}
A portion of this research was carried out at the Jet Propulsion Laboratory, California Institute of Technology, under a contract with the National Aeronautics and Space Administration. \copyright 2021 California Institute of Technology. Government sponsorship acknowledged.

\bibliographystyle{AAS_publication}   
\bibliography{references}   

\end{document}